\documentclass[10pt]{article}
\usepackage[utf8]{inputenc}
\usepackage{cite}
\usepackage{amsmath,amssymb,amsfonts}
\usepackage{algorithmic}
\usepackage{graphicx}
\usepackage{braket}
\usepackage{textcomp}
\usepackage[left=1.65cm,right=1.65cm,top=1.9cm]{geometry}
\usepackage{xcolor}
\usepackage{tikz}
\usetikzlibrary{arrows.meta, positioning, shapes.geometric, fit, backgrounds}
\usepackage{booktabs}
\def\BibTeX{{\rm B\kern-.05em{\sc i\kern-.025em b}\kern-.08em
    T\kern-.1667em\lower.7ex\hbox{E}\kern-.125emX}}
\usepackage[colorlinks=true]{hyperref}
\usepackage{url}

\newcommand{\xyzg}{\textsc{xyz\_groups}}
\newcommand{\greedy}{\textsc{greedy}}
\newcommand{\gurobi}{\textsc{gurobi}}
\newcommand{\handcrafted}{\textsc{handcrafted}}
\newcommand{\majtrn}{\textsc{MajTrain}}
\newcommand{\majreg}{\textsc{MajRegime}}

\begin{document}

% \title{Generalizing Trotter Ordering to Larger Systems with a Structure-Aware Transformer for 1D Heisenberg Hamiltonians}

% \title{A Structure-Aware Transformer for Learning Optimal Trotter Ordering Generalizes to Larger Systems for 1D Heisenberg Hamiltonians}

\title{Structure-Aware Transformers for Learning Near-Optimal Trotter Orderings with System-Size Generalization in 1D Heisenberg Hamiltonians}

% \title{Coloring-Aware Transformer for Trotter Ordering with System-Size Generalization in 1D Heisenberg Hamiltonians}

\author{Shamminuj Aktar,  Reuben Tate, Stephan Eidenbenz}

\date{
    \small
    Computing and Artificial Intelligence Division (CAI-3), \\
    Los Alamos National Laboratory, Los Alamos, New Mexico 87545, USA \\
    Email: saktar@lanl.gov, rtate@lanl.gov, eidenben@lanl.gov
}

\maketitle

\begin{abstract}
Trotterization is a standard approach for simulating quantum time evolution on quantum computers, where the Hamiltonian is split into local terms and each term is applied in sequence. The order of these terms affects the fidelity of the simulation when they do not commute, so the choice of ordering directly impacts the accuracy of the simulation.
We study this problem for one-dimensional XXZ Heisenberg Hamiltonians using a structured set of $24$ candidate orderings derived from colorings of the Hamiltonian's commutation graph and their group permutations. Finding the best candidate for large systems becomes prohibitive because fidelity evaluation is computationally expensive. In this work, we train a transformer encoder on smaller systems to predict the best candidate ordering for larger systems directly from Hamiltonian and Trotter-configuration features, without computing candidate fidelities at inference time.
The model is trained on in-range systems of $3$ to $14$ qubits with $15$-qubit systems held out for validation. Experimental results show that the model reaches a mean test fidelity gap of $0.00115$ relative to the best of the $24$ candidates on out-of-range systems of $16$ to $20$ qubits. A training-size sweep further shows that generalization emerges once training includes systems up to $L=8$ qubits, with validation at $L=9$, and the gap continues to decrease as the training range grows.
To our knowledge, this is the first application of a learned model to Trotter ordering, and it motivates future work on AI-guided Trotter ordering with generalization across Hamiltonian families and system types.
\end{abstract}

\textbf{Keywords: }
Quantum Time Evolution, Trotterization, Hamiltonian Dynamics, AI-Assisted Quantum Computing

\section{Introduction}
\label{sec:intro}

Time evolution of quantum many-body systems is a central workload for quantum computers, with applications in chemistry, materials science, and condensed matter physics~\cite{haah2018quantum, georgescu2014quantum}. Because the full time-evolution operator is generally not native to quantum hardware, digital quantum simulation approximates it with a sequence of gates. The most widely used approach is the product formula, also called Trotterization~\cite{trotter1959product, lloyd1996universal, suzuki1991general, childs2021trotter}, in which the Hamiltonian is decomposed into Pauli terms and the exponential of each term is applied in sequence. When the terms do not commute, the order of application affects the resulting unitary and therefore the simulation fidelity. This makes Trotter ordering a natural target for a learned model that can infer good choices from Hamiltonian structure and select among the structured candidates without computing their fidelities at inference time. Although worst-case Trotter error bounds are well understood, they are often far looser than the errors observed in practice~\cite{babbush2015error}.

Prior work has shown that ordering can have a large practical effect. Simulation accuracy is commonly measured by state fidelity, the squared overlap between the exact state evolved under $H$ and the approximate state produced by the Trotter formula. Simple ordering rules can differ in fidelity by an order of magnitude on realistic instances~\cite{tranter2019ordering, tate2026commutation}, and the best ordering depends on both the Hamiltonian instance and the Trotter configuration~\cite{tate2026commutation}. Choosing a good ordering is therefore a combinatorial optimization problem: a Hamiltonian with $k$ terms has $k!$ possible orderings. In principle, one could search this space by repeated classical simulation, but that approach quickly becomes impractical as the number of terms grows.

\begin{figure*}[!t]
\centering
\includegraphics[width=\linewidth]{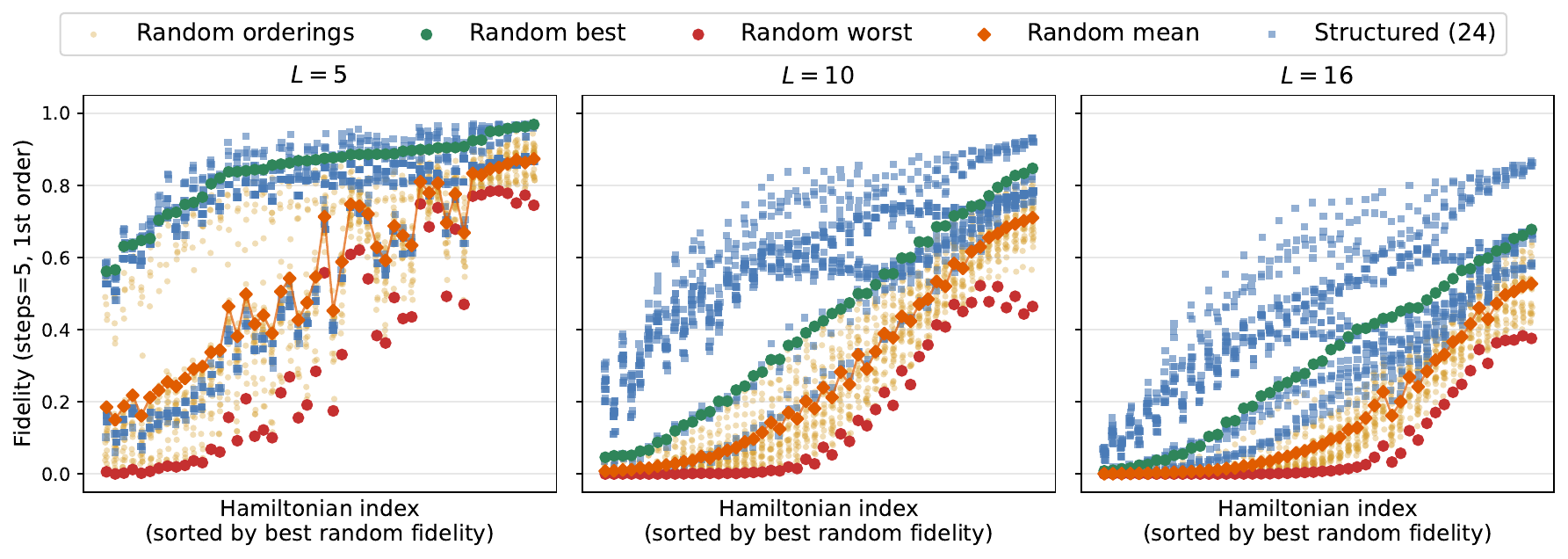}
\caption{Fidelity of random and structured Trotter orderings across system sizes $L=5, 10, 16$ for the N\'eel initial state, first-order product formula, and $r=5$ Trotter steps. For each Hamiltonian instance, sorted by best random-ordering fidelity, we show $30$ random orderings (amber dots), their best, worst, and mean fidelities (green, red, and orange), and all $24$ structured candidates (blue squares). At $L=5$, random and structured orderings are often comparable. By $L=10$, a clear gap emerges, and at $L=16$ the structured candidates consistently occupy the high-fidelity region while random orderings are concentrated lower. This trend shows that structured orderings become increasingly advantageous as system size grows.}
\label{fig:random_vs_structured}
\end{figure*}

Figure~\ref{fig:random_vs_structured} illustrates the challenge. Random orderings spread across a wide fidelity range, and that range widens with system size, while the structured candidates remain concentrated at substantially higher fidelity. The figure also suggests an important opportunity: although the full ordering space is enormous, a relatively small structured subset already captures much of the high-fidelity region. Since fidelity evaluation itself scales exponentially with system size, neither random search nor exhaustive enumeration is practical at the scales of interest. This makes Trotter ordering a natural target for a learned model that can infer good choices from Hamiltonian structure without evaluating every candidate directly.

We address this problem with, to our knowledge, the first learned approach to Trotter ordering. Machine learning has been applied to different stages of quantum compilation, including gate-level optimization~\cite{fosel2021rlcircuit}, circuit transpilation~\cite{quetschlich2023compiler}, and fidelity prediction~\cite{wang2022quest}. Trotter ordering has so far been approached using analytical and heuristic rules. Rather than search the full $k!$ space, we restrict attention to a structured candidate set derived from commutation-graph colorings and permutations of the resulting groups~\cite{tate2026commutation}. This reduction yields $24$ candidate orderings per instance in our setting. The resulting task is no longer open-ended search over all permutations, but supervised selection over a fixed set of strong candidates. Even within this reduced set, however, identifying the best candidate still requires expensive fidelity evaluation, motivating a learned selector that can generalize across system sizes.

We instantiate the approach on one-dimensional XXZ Heisenberg Hamiltonians. A transformer encoder~\cite{vaswani2017attention} trained on small systems takes the Hamiltonian terms together with the Trotter configuration and learns to predict the best candidate ordering for unseen system sizes. The input is size-invariant by construction: rather than encoding absolute system size, the model sees each Hamiltonian term individually along with a few global properties of the instance, allowing it to transfer directly to larger systems. This size-invariant formulation is central to our goal of replacing repeated per-instance simulation with a predictor that extrapolates beyond the system sizes seen during training.

This paper makes the following contributions:

\begin{itemize}

  \item We formulate Trotter ordering as a supervised classification problem over a structured candidate set, reducing the original $k!$ search space to selection among $24$ high-quality orderings.

  \item We introduce a transformer-based selector for this task and show that it generalizes beyond the system sizes seen during training.

  \item Trained on systems of $3$ to $14$ qubits, the model achieves a mean test fidelity gap of $0.00115$ on held-out systems of $16$ to $20$ qubits, more than a factor of five smaller than the strongest non-learned baseline.

  \item We report a training-range sweep showing that generalization emerges when training includes systems up to $L=8$ qubits and validation is performed at $L=9$, with continued improvement as the training range grows.

  \item We position this work as a first step toward AI-assisted Trotter ordering with generalization across Hamiltonian families, Trotter regimes, and system types.

\end{itemize}

The remainder of this paper is organized as follows. Section~\ref{sec:related} reviews related work on Trotterization, term ordering, and machine learning for quantum circuits. Section~\ref{sec:background} introduces the necessary background, and Section~\ref{sec:method} presents the proposed coloring-aware transformer approach. Section~\ref{sec:setup} describes the experimental setup, while Section~\ref{sec:results} reports the experimental results and analysis. Section~\ref{sec:discussion} discusses the limitations of the approach and outlines directions for future work. Section~\ref{sec:conclusion} concludes the paper.

\begin{figure*}[t!]
  \centering
  \includegraphics[width=\linewidth]{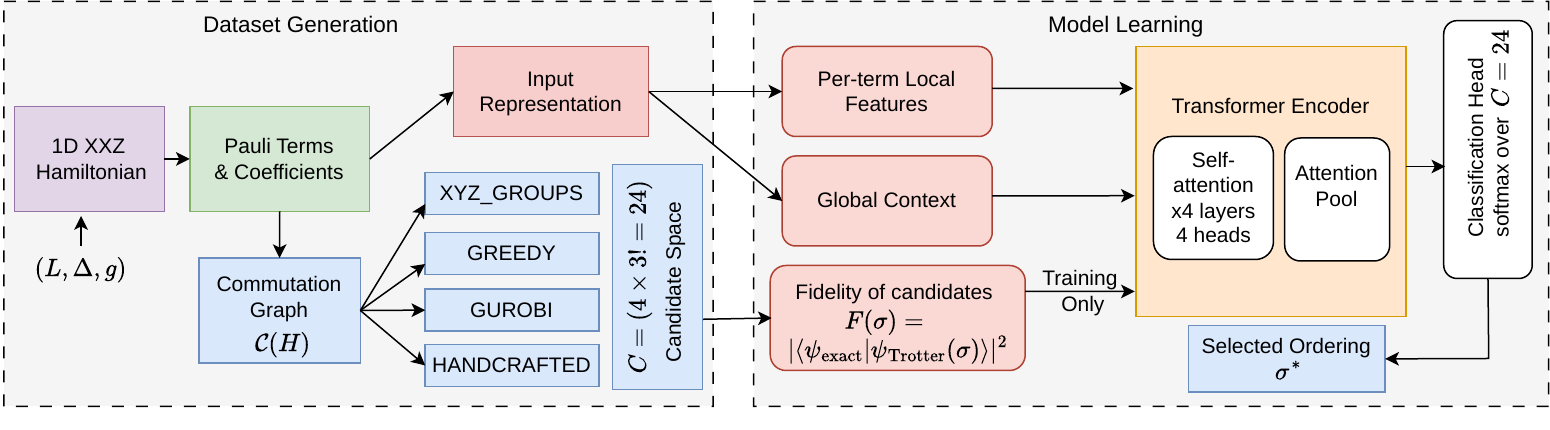}
\caption{Overview of the coloring-aware Trotter ordering pipeline, divided into two stages.
\textit{Dataset generation (left):} For a given 1D XXZ Hamiltonian $(L,\Delta,g)$, we
construct the commutation graph $\mathcal{C}(H)$ from Pauli terms and apply four coloring methods
(\xyzg, \greedy, \gurobi, \handcrafted), producing $C=24$ candidate orderings
($4\times3!=24$). For each candidate ordering, we compute the fidelity
$F(\sigma)=|\langle\psi_{\mathrm{exact}}|\psi_{\mathrm{Trotter}}(\sigma)\rangle|^2$
between the exact and Trotter-evolved states. These fidelities are used during training only:
the oracle fidelity, defined as the best fidelity among the $24$ structured candidates, provides the supervision label.
\textit{Model learning (right):} Per-term local features and global context are fed into a transformer encoder with
four self-attention layers, four heads, and attention pooling. A classification head maps the pooled representation to a softmax distribution over
$C=24$ classes; the argmax selects $\sigma^*$ at inference without evaluating candidate fidelities.}
  \label{fig:overview}
\end{figure*}

\section{Related Work}
\label{sec:related}

\subsection{Trotterization and Hamiltonian Simulation}

Product-formula methods are a foundation of digital quantum simulation. Lloyd~\cite{lloyd1996universal} established product formulas as a practical route to simulating local quantum systems on gate-based quantum computers, and Suzuki~\cite{suzuki1991general} developed the recursive constructions underlying higher-order formulas. More recent theory explains why product formulas often perform better in practice than loose worst-case bounds suggest. Childs et al.~\cite{childs2021trotter} derived commutator-scaling bounds that tie Trotter error to Hamiltonian structure, and Heyl et al.~\cite{heyl2019quantum} showed that for local observables, localization can strongly suppress Trotter errors.

Alternative simulation strategies have also been proposed. Campbell introduced qDRIFT~\cite{campbell2019random}, a randomized method that samples Hamiltonian terms according to coefficient magnitude and can outperform deterministic formulas in some regimes. Zhuk et al.~\cite{zhuk2024multiproduct} extended commutator-based analysis to multi-product formulas, showing how linear combinations of shallow product-formula circuits can improve error scaling. These approaches broaden the Hamiltonian-simulation design space.
% , but are complementary to the question studied here: for a fixed first- or second-order product formula, the ordering of Hamiltonian terms determines simulation fidelity and is the central focus of our work.

\subsection{Term Ordering and Commutation Structure}

The dependence of Trotter error on term ordering has been studied both analytically and empirically. Tranter et al.~\cite{tranter2019ordering} compared several ordering rules for molecular Hamiltonians and showed that simple choices can differ substantially in accuracy. Tate et al.~\cite{tate2026commutation} reach a similar conclusion for Heisenberg-style Hamiltonians, showing that even within a restricted structured candidate set, the best ordering varies with the Hamiltonian instance and Trotter configuration.

A common organizing principle in this literature is the commutation graph of the Hamiltonian. Gokhale et al.~\cite{gokhale2020measurement} used commuting-group structure to reduce measurement overhead in variational quantum algorithms, helping establish commutation-based partitioning as a useful way to organize Pauli terms. Tomesh et al.~\cite{li2021termgrouping} combined commuting-term grouping with traveling-salesman-style heuristics to improve simulation quality while also enabling gate cancellation. Schmitz et al.~\cite{schmitz2024graphoptimization} formulated low-depth Trotterization as a graph-optimization problem over Pauli-frame transitions, showing that graph-based search can reduce two-qubit gate counts. Our work is most directly connected to Tate et al.~\cite{tate2026commutation}, who construct structured ordering candidates from commutation-graph colorings and permutations of the resulting color classes. We adopt this candidate space and learn to select the highest-fidelity candidate for each instance.

\subsection{Machine Learning for Quantum Circuits}

Machine learning has increasingly been used in quantum compilation and circuit optimization. Fösel et al.~\cite{fosel2021rlcircuit}, Quetschlich et al.~\cite{quetschlich2023compiler}, and Moro et al.~\cite{moro2021compiling} applied reinforcement learning to problems including gate-level optimization, compiler-pass selection, and quantum compiling. Learned models have also been used to predict circuit properties without direct simulation. Wang et al.~\cite{wang2022quest} used a graph-transformer architecture to estimate the fidelity of noisy parameterized quantum circuits, replacing expensive simulation-based evaluation with a learned predictor. Aktar et al.~\cite{aktar2024gnn} showed that graph neural networks can estimate expressibility-related properties of parameterized quantum circuits with generalization beyond the qubit counts seen during training.

In this work, we learn to choose the ordering of Hamiltonian terms in a Trotterized time-evolution circuit to minimize approximation error. To our knowledge, this is the first learned method designed specifically for Trotter ordering, and the first to study generalization across system sizes using a size-invariant representation of Hamiltonian terms together with commutation-structure features.

\section{Background}
\label{sec:background}

\subsection{Trotterization and ordering error}
\label{sec:bg_trotter}

Given a Hamiltonian $H = \sum_{j=1}^{k} H_j$ written as a sum of $k$ terms, the time-evolution operator $U(t) = e^{-iHt}$ is approximated by a product formula that divides $t$ into $r$ steps and applies single-term exponentials within each step. The first-order formula is
\begin{equation}
    S_1(t/r) = \prod_{j=1}^{k} e^{-i H_j t / r},
    \label{eq:s1}
\end{equation}
with $U(t) \approx S_1(t/r)^r$. The second-order formula is
\begin{equation}
    S_2(t/r) = \prod_{j=1}^{k} e^{-i H_j t / 2r} \prod_{j=k}^{1} e^{-i H_j t / 2r},
    \label{eq:s2}
\end{equation}
with $U(t) \approx S_2(t/r)^r$, where the reversed second pass cancels the leading first-order error term, and higher orders follow the Suzuki recursion~\cite{suzuki1991general}. For a $p$-th-order product formula, the approximation error typically scales as $\mathcal{O}(t^{p+1}/r^p)$, with a prefactor controlled by nested commutators of the $H_j$~\cite{childs2021trotter}. Since $[H_i, H_j] \ne 0$ in general, the product in Eq.~\ref{eq:s1} depends on the order in which terms are applied, and different orderings yield different approximations.

% ----------------------------------------------------------------
%  Table 1 – Color-group assignments (tab:coloring_groups)
%  Fix: wrap long cell content, reduce font, use p{} columns
% ----------------------------------------------------------------
\begin{table*}[t!]
  \centering
  \caption{Color-group assignments for the 13 Pauli terms of the $L=4$ XXZ Hamiltonian under the four
    coloring methods \xyzg, \greedy, \gurobi, and \handcrafted, denoted by $m_1, m_2, m_3, m_4$,
    respectively. Each method partitions the terms into three color classes such that terms within a
    class mutually commute.}
  \label{tab:coloring_groups}
  \small
  \setlength{\tabcolsep}{4pt}
  \begin{tabular}{@{}l l p{4.8cm} p{4.8cm} p{3.2cm}@{}}
    \toprule
    Method              & Index & Group 0 & Group 1 & Group 2 \\
    \midrule
    \xyzg        & $m_1$ & $X_1X_2, X_2X_3, X_3X_4, X_1, X_2, X_3, X_4$
                         & $Y_1Y_2, Y_2Y_3, Y_3Y_4$
                         & $Z_1Z_2, Z_2Z_3, Z_3Z_4$ \\[2pt]
    \greedy      & $m_2$ & $X_2X_3, Y_2Y_3, Z_2Z_3, X_1, X_4$
                         & $X_1X_2, X_3X_4, Y_1Y_2, Y_3Y_4, Z_1Z_2, Z_3Z_4$
                         & $X_2, X_3$ \\[2pt]
    \gurobi      & $m_3$ & $X_1X_2, X_3X_4, Y_1Y_2, Y_3Y_4, Z_1Z_2, Z_3Z_4$
                         & $X_2X_3, Y_2Y_3, Z_2Z_3, X_1$
                         & $X_2, X_3, X_4$ \\[2pt]
    \handcrafted & $m_4$ & $X_1X_2, X_3X_4, Y_1Y_2, Y_3Y_4, Z_1Z_2, Z_3Z_4$
                         & $X_2X_3, Y_2Y_3, Z_2Z_3$
                         & $X_1, X_2, X_3, X_4$ \\
    \bottomrule
  \end{tabular}
\end{table*}

\subsection{1D XXZ Heisenberg Hamiltonian}
\label{sec:bg_xxz}

We study the one-dimensional spin-$1/2$ XXZ Heisenberg model with an external transverse field:
\begin{equation}
    H = \sum_{i=1}^{L-1} \left(X_i X_{i+1} + Y_i Y_{i+1} + \Delta\, Z_i Z_{i+1}\right) + g \sum_{i=1}^{L} X_i,
    \label{eq:xxz}
\end{equation}
with open boundaries, $L$ qubits, anisotropy $\Delta$, and transverse field $g$~\cite{auerbach2012interacting}. The Hamiltonian contains $4L-3$ Pauli terms.

\subsection{Commutation-graph colorings}
\label{sec:bg_coloring}

Following Tate et al.~\cite{tate2026commutation}, we structure the ordering problem using the commutation graph of the Hamiltonian. The graph has one node for each term in $H$, and an edge connects two nodes when the corresponding terms do not commute. Finding commuting groups is then an instance of the classical graph-theory problem of vertex coloring: partitioning the vertices into independent sets, or color classes, such that no two vertices within the same class share an edge. By construction, each color class corresponds to a group of mutually commuting Hamiltonian terms, so terms within a class can be applied in any internal order.

For a coloring with $n_c$ color classes, this reduces the structured search space from $k!$ full term permutations to $n_c!$ color-class permutations, where typically $n_c \ll k$. Different coloring methods can produce different partitions and therefore different candidate orderings. For the 1D XXZ Hamiltonian of Eq.~\ref{eq:xxz}, the colorings used in this work always produce three color classes, independent of system size $L$. Enumerating all $3! = 6$ permutations of these classes for each of the four coloring methods from Tate et al.~\cite{tate2026commutation} gives $4 \times 3! = 24$ structured candidate orderings, which form the candidate space for the transformer encoder described in Section~\ref{sec:method}.

\section{Method}
\label{sec:method}

We treat Trotter ordering as a multi-class classification problem over a structured set of $24$ candidate orderings. This section defines the task, the candidate space, the input representation, and the model. Figure~\ref{fig:overview} shows the full pipeline. At a high level, the pipeline first constructs a structured set of $24$ ordering candidates from commutation-graph colorings, then trains a transformer to select the best candidate from Hamiltonian features and Trotter configuration. Fidelity values are used only during dataset generation and training to define the supervision label; at inference, the model receives only the Hamiltonian features and Trotter configuration and outputs one of the $24$ candidate classes directly.

\subsection{Problem formulation}
\label{sec:task}

Let $H = \sum_{j=1}^{k} c_j P_j$ be a Hamiltonian written as a weighted sum of Pauli-string terms, with real coefficients $c_j$ and local Pauli strings $P_j$.

For a given Trotter order $p \in \{1,2\}$ and step count $r \in \{3,5\}$, a product formula approximates $e^{-iHt}$ by repeating an ordering-dependent single-step formula $r$ times. For first-order Trotter, an ordering $\sigma$ gives
\begin{equation}
    S_{1,\sigma}(t/r)
    =
    \prod_{j=1}^{k}
    e^{-i c_{\sigma(j)} P_{\sigma(j)} t/r}.
\end{equation}
For second-order Trotter, the corresponding step uses the same ordering followed by its reverse, as in Eq.~\ref{eq:s2}. The approximation error depends on $\sigma$ in both cases. We consider a fixed target time $t=5$ and a fixed initial state $|\psi_0\rangle$ (the N\'eel state $|0101\ldots\rangle$~\cite{auerbach2012interacting}). Simulation fidelity is measured as
\begin{equation}
    F(\sigma)
    =
    \left|
    \left\langle
    \psi_{\mathrm{exact}}
    \middle|
    \psi_{\mathrm{Trotter}}(\sigma)
    \right\rangle
    \right|^2,
\end{equation}
where
$|\psi_{\mathrm{exact}}\rangle = e^{-iHt}|\psi_0\rangle$
is the exact evolved state and
$|\psi_{\mathrm{Trotter}}(\sigma)\rangle
= U_{\mathrm{Trotter}}(\sigma)|\psi_0\rangle$
is the approximate state produced by ordering $\sigma$~\cite{jozsa1994fidelity}.

Given an instance $(H,p,r)$, our goal is to predict the candidate $\sigma^*$ that maximizes $F$, while avoiding any fidelity computation over the candidate set at inference time. Rather than predicting fidelity directly, the model treats the problem as selection among a finite set of $C$ candidate orderings and is trained with cross-entropy loss over the $C$ classes. Full training details are given in Section~\ref{sec:training_setup}.

\subsection{Candidate space}
\label{sec:candidates}

We use $C=24$ candidate orderings per instance, following the group-evolve construction of Tate et al.~\cite{tate2026commutation}. For the 1D XXZ commutation graph we apply four coloring methods, namely Pauli-type grouping (\xyzg), a greedy heuristic (\greedy), an exact minimum coloring via integer programming (\gurobi), and a handcrafted bond-parity coloring (\handcrafted). Under our 1D XXZ setup, each method produces three color classes for the Hamiltonian of Eq.~\ref{eq:xxz}. The \xyzg\ coloring yields three classes because all three Pauli types $\{X, Y, Z\}$ are present. The other three methods also yield three classes because the bond-parity argument splits even-bond and odd-bond two-body terms into two classes, with the single-site $X_i$ terms occupying the third. Enumerating the $3! = 6$ permutations of the three classes for each of the four colorings gives $4 \times 6 = 24$ orderings. This set is small enough for supervised learning while still containing strong candidates for the instances we study.

% This set is small enough for supervised learning while still containing a near-optimal ordering for most instances we study.

\begin{table}[t!]
  \centering
  \caption{Per-term feature encoding for the $13$ Pauli-string terms of the $L=4$ XXZ Hamiltonian
    in Table~\ref{tab:coloring_groups} ($\Delta=0.25$, $g=1.8$). The $\log|c|$ column uses
    file-stored coefficients under the spin-$\tfrac{1}{2}$ convention ($S^\alpha = \tfrac{1}{2}\sigma^\alpha$),
    so two-body terms carry a factor of $\tfrac{1}{4}$ and single-site terms a factor of $\tfrac{1}{2}$
    relative to the bare Pauli operators. The Pauli column is a six-way one-hot
    ($X$, $Y$, $Z$, $XX$, $YY$, $ZZ$); body indicates whether the term is single-qubit ($0$) or
    two-body ($1$); dist is the inter-qubit distance; $m_1, m_2, m_3, m_4$ are the color-group
    indices from Table~\ref{tab:coloring_groups}. The global context shared across all terms is
    shown below the rule.}
  \label{tab:features}
  \scriptsize
  \resizebox{0.9\textwidth}{!}{%
  \begin{tabular}{lcccccccc}
    \toprule
    Term         & $\log|c|$ & Pauli           & body & dist & $m_1$ & $m_2$ & $m_3$ & $m_4$ \\
    \midrule
    $X_1 X_2$    & $-1.386$ & $[0,0,0,1,0,0]$ & $1$ & $1$ & $0$ & $1$ & $0$ & $0$ \\
    $Y_1 Y_2$    & $-1.386$ & $[0,0,0,0,1,0]$ & $1$ & $1$ & $1$ & $1$ & $0$ & $0$ \\
    $Z_1 Z_2$    & $-2.773$ & $[0,0,0,0,0,1]$ & $1$ & $1$ & $2$ & $1$ & $0$ & $0$ \\
    $X_2 X_3$    & $-1.386$ & $[0,0,0,1,0,0]$ & $1$ & $1$ & $0$ & $0$ & $1$ & $1$ \\
    $Y_2 Y_3$    & $-1.386$ & $[0,0,0,0,1,0]$ & $1$ & $1$ & $1$ & $0$ & $1$ & $1$ \\
    $Z_2 Z_3$    & $-2.773$ & $[0,0,0,0,0,1]$ & $1$ & $1$ & $2$ & $0$ & $1$ & $1$ \\
    $X_3 X_4$    & $-1.386$ & $[0,0,0,1,0,0]$ & $1$ & $1$ & $0$ & $1$ & $0$ & $0$ \\
    $Y_3 Y_4$    & $-1.386$ & $[0,0,0,0,1,0]$ & $1$ & $1$ & $1$ & $1$ & $0$ & $0$ \\
    $Z_3 Z_4$    & $-2.773$ & $[0,0,0,0,0,1]$ & $1$ & $1$ & $2$ & $1$ & $0$ & $0$ \\
    $X_1$        & $-0.105$ & $[1,0,0,0,0,0]$ & $0$ & $0$ & $0$ & $0$ & $1$ & $2$ \\
    $X_2$        & $-0.105$ & $[1,0,0,0,0,0]$ & $0$ & $0$ & $0$ & $2$ & $2$ & $2$ \\
    $X_3$        & $-0.105$ & $[1,0,0,0,0,0]$ & $0$ & $0$ & $0$ & $2$ & $2$ & $2$ \\
    $X_4$        & $-0.105$ & $[1,0,0,0,0,0]$ & $0$ & $0$ & $0$ & $0$ & $2$ & $2$ \\
    \midrule
    \midrule
    \multicolumn{9}{l}{\textit{Global context (shared across all terms)}} \\[4pt]
    \midrule
    \multicolumn{2}{c}{$p$} &
    \multicolumn{2}{c}{$r$} &
    \multicolumn{2}{c}{$\sigma_{|c|}/\max|c|$} &
    \multicolumn{1}{c}{$\sum|c_{ZZ}|/\sum|c_X|$} &
    \multicolumn{2}{c}{Two-body fraction} \\[2pt]
    \midrule
    \multicolumn{2}{c}{$\in\{1,2\}$} &
    \multicolumn{2}{c}{$\in\{3,5\}$} &
    \multicolumn{2}{c}{$0.374$} &
    \multicolumn{1}{c}{$0.052$} &
    \multicolumn{2}{c}{$0.692\ (9/13)$} \\
    \bottomrule
  \end{tabular}}
\end{table}

\subsection{Input representation}
\label{sec:input}

Each sample is encoded using the Pauli terms and coefficients of the Hamiltonian, represented as a set of per-term features together with a global context vector. This representation is size-invariant in the sense that it does not depend on absolute qubit indices or a fixed number of Hamiltonian terms, which enables evaluation at system sizes beyond those seen in training.

\paragraph{Per-term features.} For each term $j$ we encode:
\begin{itemize}
  \item the log-magnitude of the coefficient, $\log|c_j|$,
  \item a one-hot encoding over a six-entry Pauli-type vocabulary ($X$, $Y$, $Z$, $XX$, $YY$, $ZZ$), of which four ($X$, $XX$, $YY$, $ZZ$) are populated by the 1D XXZ Hamiltonian,
  \item the term body ($0$ for a single-qubit term, $1$ for a two-body term),
  \item the inter-qubit distance of the term ($0$ for a single-qubit term, $1$ for a nearest-neighbor two-body term),
  \item four color-group indices $m_1, m_2, m_3, m_4$, one per method (\xyzg, \greedy, \gurobi, \handcrafted), giving the integer color group of the term under that method.
\end{itemize}

\paragraph{Global context.} The model also receives the Trotter order $p$, the step count $r$, and three scale-invariant global statistics of the Hamiltonian coefficients, namely (i) the standard deviation over the maximum coefficient, (ii) the $ZZ$-to-$X$ coefficient magnitude ratio (approximately $\Delta/g$ for XXZ), and (iii) the fraction of two-body terms.

We deliberately exclude absolute qubit indices from the input. The model has no way to tell whether a term acts on qubit $3$ or qubit $17$, which forces it to learn a decision rule based on term statistics and structure, and enables evaluation at larger system sizes.

Table~\ref{tab:coloring_groups} illustrates how the four coloring methods differ on a concrete $L=4$ XXZ instance. \xyzg\ separates by Pauli type, \handcrafted\ separates by bond parity with single-$X$ terms isolated, \gurobi\ is similar but groups $X_1$ with the inner bond, and \greedy\ produces a mixed partition. The four methods therefore offer four distinct structural views of the same Hamiltonian, and the model learns to weight them via the per-term color-group indices. Table~\ref{tab:features} shows how the $13$ Pauli terms of the $L=4$ Hamiltonian are encoded as per-term feature vectors, along with the global context features shared across all terms.

\subsection{Model architecture}
\label{sec:model}

We use a transformer encoder over the unordered set of term features~\cite{vaswani2017attention}. Each term's categorical features (operator type, term body, inter-qubit distance, and the four color-group indices) are mapped to learned embeddings, concatenated with the log coefficient $\log|c_j|$, and projected to $d_\text{model} = 128$. A stack of four pre-norm transformer encoder blocks ($4$ heads, $d_\text{ff} = 256$, dropout $0.1$) applies multi-head self-attention across the term set. Because we omit positional encoding, self-attention is permutation-equivariant on the term set. The encoder output is pooled into a single summary vector by attention-based pooling, where a two-layer scoring network assigns a scalar score to each term. Scores are softmax-normalized over non-padded positions, and the term embeddings are combined as the resulting weighted sum. The pooled summary is concatenated with the global context (embedded Trotter order, step count, and the three scale-invariant statistics), passed through a projection head to a $128$-dimensional hidden state, and mapped by a linear head to the $C = 24$ class logits.

The full model has $662{,}421$ trainable parameters, with the transformer stack accounting for roughly $80\%$. We chose this architecture because the input is an unordered set of term descriptors, which self-attention without positional encoding handles naturally.

\section{Experimental Setup}
\label{sec:setup}

\subsection{Dataset}
\label{sec:dataset}

We generate a dataset of Trotter-simulation fidelities for the 1D XXZ Hamiltonian of Eq.~\ref{eq:xxz}. System size ranges from $L=3$ to $L=20$ qubits. For each $L$ we sweep two anisotropy values $\Delta \in \{0.12, 0.25\}$ and twenty-five transverse-field values $g \in \{0.1, 0.2, \ldots, 2.5\}$, giving $50$ Hamiltonian instances per $L$ and $900$ instances across the full range. For each instance, during dataset generation, we evaluate all $C=24$ structured candidate orderings and $30$ random orderings at Trotter orders $p \in \{1, 2\}$ and step counts $r \in \{3, 5\}$. The target time is fixed to $t=5$ and the initial state is the N\'eel state $|0101\ldots\rangle$. For each structured candidate ordering, we compute the fidelity between the exact and Trotter-evolved states by exact classical simulation using matrix exponentiation. A sample is a single $(L, \Delta, g, p, r)$ configuration together with the fidelities of its $24$ structured candidates, yielding $3{,}600$ samples in total. These candidate fidelities are used to define the oracle class label and to evaluate test metrics, but they are not used by the model at inference time. For the main experiment, we split samples by system size to evaluate extrapolation to larger systems. The training set uses $L \in \{3,4,\ldots,14\}$, the validation set is $L=15$, and the test set is $L \in \{16,17,18,19,20\}$, all strictly larger than any $L$ seen during training. Within each $L$, no further splitting is applied, so the model sees every $(\Delta, g, p, r)$ combination for $L$ in the training range.

\subsection{Baselines}
\label{sec:baselines}

We compare the learned selector against three non-learned baselines of increasing sophistication. \emph{Random} uses the mean fidelity over $30$ random orderings per instance and represents the zero-information baseline. \emph{MajTrain} picks the single candidate class that wins most often across the entire training set and applies it to every test instance unconditionally, giving the strongest fixed policy that does not condition on the input instance. \emph{MajRegime} picks the class that wins most often in the training subset sharing the same Trotter order $p$ and step count $r$ as the test instance. It is the strongest non-learned policy because it conditions on the same regime variables $(p, r)$ that the model sees, but ignores all per-instance Hamiltonian information. \majreg\ is the key baseline to beat, since any improvement over it must come from per-instance information beyond the regime variables $(p, r)$.

\subsection{Metrics}
\label{sec:metrics}

Let $f_j^*$ be the oracle fidelity, defined as the best fidelity among the $24$ structured candidates for test instance $j$, and let $f_j^{\text{model}}$ be the fidelity of the candidate selected by the model. We report three quantities. The \emph{test fidelity gap}, defined as $f_j^* - f_j^{\text{model}}$ and averaged over test instances, is the headline metric and is zero when the model selects the oracle candidate. The \emph{normalized gain}, defined per instance as
\[
(f_j^{\text{model}} - f_j^{\text{rand}})/(f_j^* - f_j^{\text{rand}})
\]
and then averaged over valid test instances with $f_j^* > f_j^{\text{rand}}$, measures how much of the random-to-oracle improvement the model captures, with values near $1.0$ indicating oracle-level performance. The \emph{poor-prediction rate}, defined as the fraction of test instances with test fidelity gap above $0.01$, highlights tail behavior. Because the oracle $f_j^*$ is defined over the $24$ structured candidates only, the random-ordering mean $f_j^{\mathrm{rand}}$ can occasionally exceed $f_j^*$ on individual instances. We exclude these cases from the normalized-gain average and interpret normalized gain as a relative baseline-improvement statistic rather than as a bounded accuracy metric.

% ----------------------------------------------------------------
%  Table 3 – Main results (tab:main)
%  Fix: was 12 data columns + row-label column = 13 cols total,
%  causing overflow on a two-column layout.
%  Solution: \resizebox to \textwidth, \small font, tighter
%  \tabcolsep, and abbreviated column headers.
% ----------------------------------------------------------------
\begin{table*}[t!]
  \centering
  \caption{Per-regime breakdown of the three evaluation metrics, averaged over five seeds and all
    held-out test sizes $L \in \{16,\ldots,20\}$. \emph{Gap}: test fidelity gap
    $f^* - f^{\mathrm{model}}$, averaged over test instances (lower~$\downarrow$ is better).
    \emph{Norm}: normalized gain $(f^{\mathrm{model}} - f^{\mathrm{rand}})/(f^* - f^{\mathrm{rand}})$
    on instances where $f^* > f^{\mathrm{rand}}$ (higher~$\uparrow$ is better).
    \emph{Poor}: fraction of instances with gap $> 0.01$ (lower~$\downarrow$ is better).
    All model metrics are computed without evaluating candidate fidelities at inference time.
    The model performs near-oracle under first-order Trotter across both step counts.
    Errors concentrate in the second-order regime (poor-prediction rate up to $6.0\%$);
    even so, the model achieves normalized gain close to $1.0$ and substantially improves
    over \majreg\ in overall mean gap, with the largest improvement in the hardest
    second-order $r{=}3$ regime.}
  \label{tab:main}
  \resizebox{\textwidth}{!}{%
  \setlength{\tabcolsep}{5pt}
  \begin{tabular}{l cccc cccc cccc}
    \toprule
    & \multicolumn{6}{c}{\textbf{1st-order Trotter}}
    & \multicolumn{6}{c}{\textbf{2nd-order Trotter}} \\
    \cmidrule(lr){2-7} \cmidrule(lr){8-13}
    & \multicolumn{3}{c}{$r=3$} & \multicolumn{3}{c}{$r=5$}
    & \multicolumn{3}{c}{$r=3$} & \multicolumn{3}{c}{$r=5$} \\
    \cmidrule(lr){2-4} \cmidrule(lr){5-7}
    \cmidrule(lr){8-10} \cmidrule(lr){11-13}
    & Gap $\downarrow$ & Norm $\uparrow$ & Poor $\downarrow$
    & Gap $\downarrow$ & Norm $\uparrow$ & Poor $\downarrow$
    & Gap $\downarrow$ & Norm $\uparrow$ & Poor $\downarrow$
    & Gap $\downarrow$ & Norm $\uparrow$ & Poor $\downarrow$ \\
    \midrule
    Random
      & $0.1754$ & $0.000$ & $100.0\%$
      & $0.4441$ & $0.000$ & $100.0\%$
      & $0.1558$ & $0.000$ & $93.0\%$
      & $0.2272$ & $0.000$ & $88.0\%$ \\
    \majtrn
      & $0.0011$ & $0.997$ & $6.0\%$
      & $0.0004$ & $0.997$ & $1.0\%$
      & $0.0957$ & $0.067$ & $91.0\%$
      & $0.1534$ & $0.268$ & $86.0\%$ \\
    \majreg
      & $0.0011$ & $0.997$ & $6.0\%$
      & $0.0004$ & $0.997$ & $1.0\%$
      & $0.0214$ & $0.786$ & $34.0\%$
      & $0.0011$ & $0.959$ & $4.0\%$ \\
    \textbf{Model (ours)}
      & $\mathbf{0.00030}$ & $\mathbf{0.999}$ & $\mathbf{0.7\%}$
      & $\mathbf{0.00052}$ & $\mathbf{0.996}$ & $\mathbf{1.8\%}$
      & $\mathbf{0.00250}$ & $\mathbf{0.973}$ & $\mathbf{6.0\%}$
      & $\mathbf{0.00127}$ & $\mathbf{0.986}$ & $\mathbf{6.0\%}$ \\
    \bottomrule
  \end{tabular}}
\end{table*}

\subsection{Training}
\label{sec:training_setup}

We train the model in Section~\ref{sec:model} as a supervised $24$-class classifier. Each training sample corresponds to one configuration $(H,p,r)$. The target label is the candidate ordering with the highest simulated fidelity among the $24$ structured candidates. We call this candidate the oracle class for that sample. The transformer produces one logit for each candidate ordering, and the cross-entropy objective encourages the oracle class to receive the largest predicted probability. Training is performed with AdamW using an initial learning rate of $3 \times 10^{-4}$ and weight decay $10^{-4}$. The learning rate is annealed with a cosine schedule over a $200$-epoch budget, reaching $3 \times 10^{-6}$ at the end of training. Because some candidate orderings win more often than others, the training labels are class-imbalanced. We therefore use class-weighted cross-entropy. Let $m_c$ be the number of training samples for which candidate class $c$ is oracle. For every class that appears in the training set, we assign the weight
\begin{equation}
    w_c =
    \frac{m_c^{-1/2}}
    {\frac{1}{|\mathcal{A}|}\sum_{c'\in\mathcal{A}} m_{c'}^{-1/2}},
    \qquad
    \mathcal{A}=\{c:m_c>0\}.
\end{equation}
This normalization keeps the average weight over observed classes equal to one, while giving larger weights to rarer oracle classes. In effect, mistakes on rare winning candidates contribute more to the loss than mistakes on candidates that win frequently. We also use label smoothing with parameter $0.05$, which reduces overconfidence in cases where several candidates have similar fidelities.

Model selection is based on validation fidelity gap rather than validation cross-entropy. This choice reflects the goal of the task: a class prediction can be wrong but still select an ordering whose fidelity is nearly oracle-level, whereas another wrong prediction may incur a much larger fidelity loss. Therefore, validation gap is more directly aligned with the reported metric than classification loss alone. We begin checkpointing after a $20$-epoch warmup and use early stopping with patience $50$ based on validation loss. The main results are averaged over five random seeds. For the generalization sweep in Section~\ref{sec:data_eff} and the sample-efficiency sweep in Section~\ref{sec:sample_eff}, we use three seeds per configuration. All models are implemented in PyTorch and trained on an Apple M-series GPU.

\section{Experimental Results}
\label{sec:results}

We present results in four parts. First, we report overall test performance of the learned selector against baselines (Section~\ref{sec:main_result}) and analyze the regimes in which errors occur (Section~\ref{sec:per_regime}). We then study generalization across system sizes (Section~\ref{sec:data_eff}) and sample efficiency under reduced numbers of Hamiltonians per system size (Section~\ref{sec:sample_eff}).

\subsection{Main result}
\label{sec:main_result}

\begin{figure}[b!]
  \centering
  \includegraphics[width=0.6\linewidth]{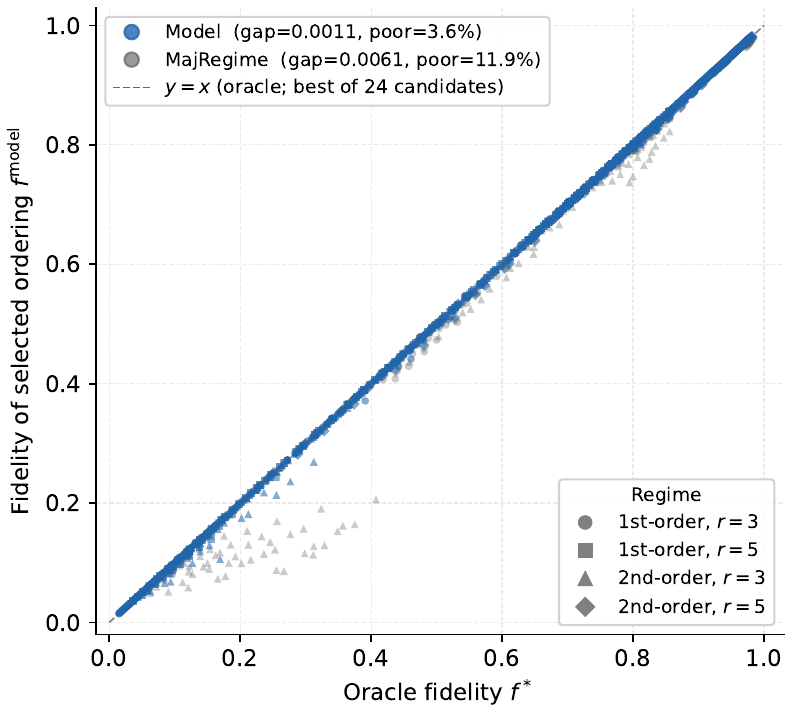}
  \caption{Selected-ordering fidelity versus oracle fidelity for the learned model and the strongest non-learned baseline, \majreg, across held-out test instances ($L \in \{16,17,18,19,20\}$). Points are colored by method and shaped by Trotter regime $(p,r)$, and the dashed line marks oracle-level performance $y=x$. The learned selector remains more tightly concentrated near the oracle line overall, while \majreg\ shows substantially larger deviations, especially in the lower-fidelity regime.}
  \label{fig:oracle_scatter}
\end{figure}

Table~\ref{tab:main} provides a full per-regime breakdown of all three evaluation metrics from Section~\ref{sec:metrics}. Across five random seeds, the learned selector achieves a mean test fidelity gap of $0.00115 \pm 0.00030$ relative to the best of the $24$ structured candidates, compared to $0.06328$ for \majtrn\ and $0.00613$ for \majreg. This is more than five times smaller than \majreg, the strongest 
non-learned baseline, and more than fifty times smaller than \majtrn. In terms of normalized gain, computed on instances where the structured oracle exceeds the random baseline, the model remains close to oracle-level performance in every regime, with the lowest regime-level value equal to $0.973$. The poor-prediction rate remains below $2\%$ in the first-order regimes and rises to at most $6\%$ in the second-order regimes. Most of the remaining mean gap is driven by a small tail of harder second-order instances, which we analyze in Section~\ref{sec:per_regime}. Figure~\ref{fig:oracle_scatter} provides a complementary view of this improvement: overall, the learned selector tracks oracle fidelity much more closely than \majreg, with the largest deviations concentrated in the more difficult
low-fidelity region.

\subsection{Error analysis}
\label{sec:per_regime}
 
\begin{figure*}[t!]
  \centering
  \includegraphics[width=0.9\linewidth]{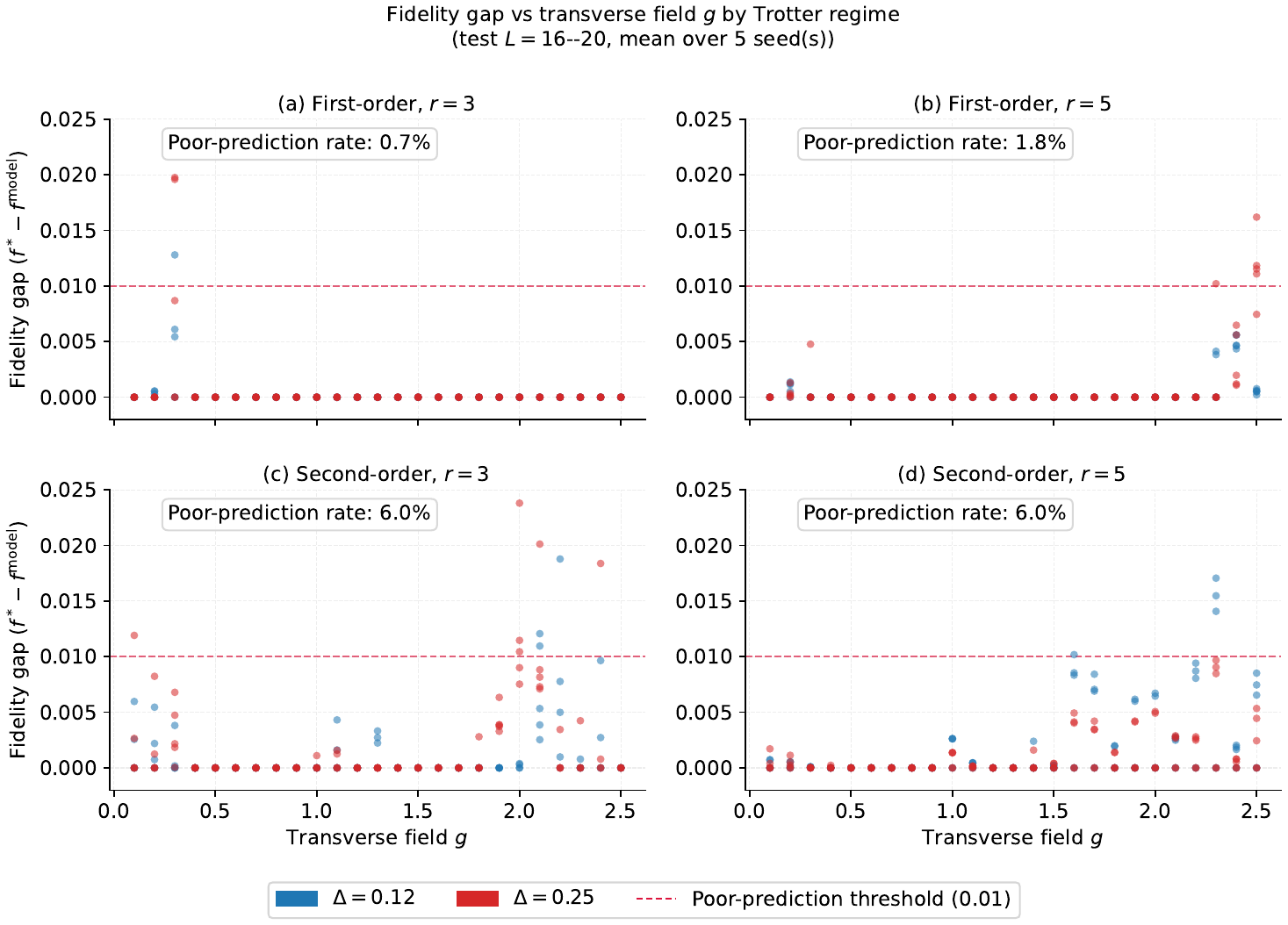}
  \caption{Error landscape: fidelity gap $f^* - f^{\mathrm{model}}$ versus transverse field $g$ for each Trotter regime, averaged over five seeds on test sizes $L \in \{16,17,18,19,20\}$. Color indicates anisotropy $\Delta$, and the dashed line marks the poor-prediction threshold (gap $= 0.01$). Under first-order Trotter (top row), the poor-prediction rate remains low at $0.7\%$ for $r=3$ and $1.8\%$ for $r=5$. Under second-order Trotter (bottom row), failures are more frequent, with poor-prediction rates of $6.0\%$ for $r=3$ and $6.0\%$ for $r=5$. The largest gaps concentrate at $g \gtrsim 2.0$, where oracle fidelity is already low and differences between candidate orderings become more pronounced.}
  \label{fig:error_landscape}
\end{figure*}

As seen in Table~\ref{tab:main}, errors concentrate in the second-order Trotter regime. Figure~\ref{fig:error_landscape} makes the structure of these errors explicit by plotting the per-instance fidelity gap $f^* - f^{\mathrm{model}}$ against the transverse field $g$ for each of the four Trotter regimes.
 
Under first-order Trotter (panels a and b), the poor-prediction rate remains low at $0.7\%$ for $r=3$ and $1.8\%$ for $r=5$, and nearly all test instances sit at or near zero gap. Under second-order Trotter (panels c and d), failures emerge more visibly, with poor-prediction rates of $6.0\%$ for both $r=3$ and $r=5$. The largest gaps appear at $g \gtrsim 2.0$, where the Hamiltonian is dominated by the transverse field and oracle fidelity is already low (around $0.1$--$0.2$). In this regime, relatively small differences between candidate orderings translate into visible fidelity gaps. The two anisotropy values show similar error patterns, indicating that the difficulty is driven primarily by $g$ rather than by the $ZZ$ coupling strength $\Delta$.
 
The training set covers all $g$ values uniformly, so the elevated poor-prediction rate at large $g$ under second-order Trotter reflects an intrinsic difficulty of the regime rather than a data imbalance. The worst per-sample gap across all five seeds is $0.057$, on an $L=16$ instance with $\Delta=0.25$, $g=1.9$, second-order Trotter at $r=3$, where the oracle fidelity is $0.156$ and the model-selected ordering reaches $0.100$. Even in this hardest case, the model retains most of the random-to-oracle improvement. Overall, the mean gap is driven by a small tail of difficult second-order instances, while the bulk of the test set remains at or near oracle fidelity.

\subsection{Generalization across system sizes}
\label{sec:data_eff}

\begin{figure}[t!]
  \centering
  \includegraphics[width=0.6\linewidth]{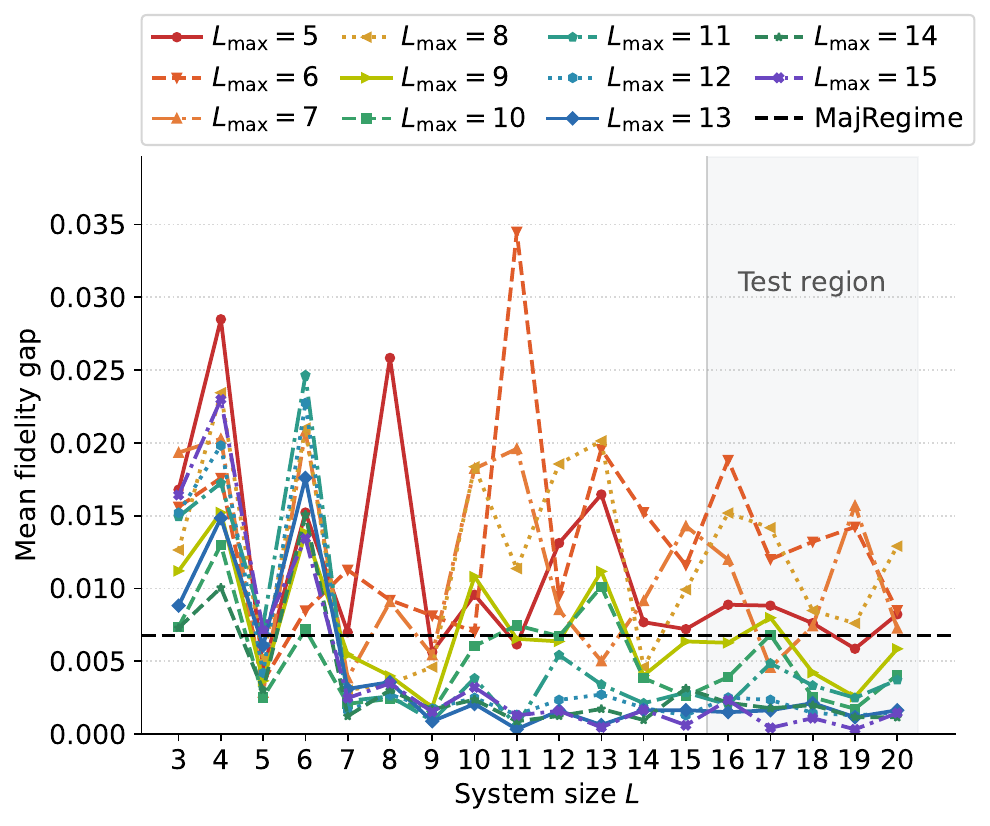}
\caption{Generalization sweep. Mean fidelity gap versus system size $L$ for each validation size $L_{\max} \in \{5,\ldots,15\}$, where training uses $L \in \{3,\ldots,L_{\max}-1\}$ and evaluation is performed on larger sizes up to $L=20$ (one line per $L_{\max}$, mean over three seeds). The shaded region marks the final held-out test range $L \in \{16,\ldots,20\}$, and the dashed horizontal line marks the \majreg\ reference. Lines for $L_{\max}\leq 8$ remain near $10^{-2}$ on larger systems, indicating limited generalization. Generalization emerges at $L_{\max}=9$, and lines for $L_{\max}\geq 13$ compress toward $10^{-3}$ with little growth across system size, confirming clean extrapolation to larger systems.}
  \label{fig:data_eff}
\end{figure}

To identify the minimum training range needed for generalization, we sweep
$L_{\max} \in \{5,6,\ldots,15\}$, training on
$L \in \{3,\ldots,L_{\max}-1\}$ and validating on $L=L_{\max}$.
For each sweep point, we evaluate on all larger system sizes
$L \in \{L_{\max}+1,\ldots,20\}$. The shaded region in
Figure~\ref{fig:data_eff} marks the final held-out test range
$L \in \{16,17,18,19,20\}$.

Figure~\ref{fig:data_eff} shows the mean fidelity gap as a function of system
size for each value of $L_{\max}$. For $L_{\max} \leq 8$, the model fails to
generalize reliably, with gaps often remaining around $10^{-2}$ and the
selector underperforming \majreg\ ($0.00613$) on larger systems. Generalization emerges at $L_{\max}=9$, corresponding to training on
$L \in \{3,\ldots,8\}$ and validation on $L=9$, where the model begins to beat \majreg\ across the larger evaluated sizes shown in Figure~\ref{fig:data_eff}. For $L_{\max} \geq 13$, the curves compress toward the bottom of the plot near $10^{-3}$, and additional training range yields only marginal further improvement. These results show that once the training range is sufficiently broad, the gap remains stable as system size increases, including throughout the final held-out test region
$L \in \{16,\ldots,20\}$.

\subsection{Sample efficiency}
\label{sec:sample_eff}
 
The generalization sweep in Section~\ref{sec:data_eff} fixes the number of Hamiltonians per system size at $50$ and varies the training range. A complementary question is how many Hamiltonians per system size are needed when the training range is fixed at $L \in \{3,\ldots,14\}$. We address this by subsampling $n_{\mathrm{ham}} \in \{10, 20, 30, 40, 50\}$ Hamiltonians per $L$ during training while keeping the full validation and test sets unchanged, and running three seeds per configuration.
 
\begin{figure}
  \centering
  \includegraphics[width=0.6\linewidth]{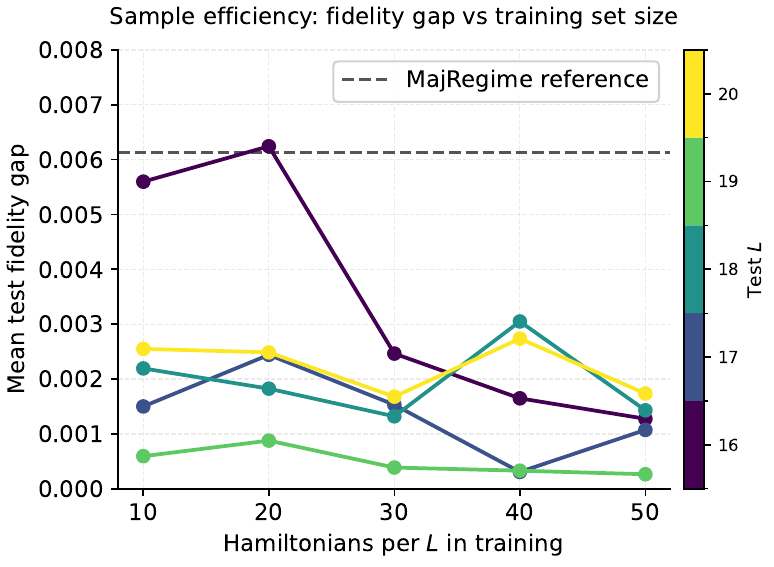}
  \caption{Sample efficiency: mean test fidelity gap versus the number of Hamiltonians per system size used in training. Colored lines show held-out test sizes $L \in \{16,17,18,19,20\}$, and the dashed line marks the average \majreg\ reference. By $n_{\mathrm{ham}}=30$, all plotted test sizes fall below the \majreg\ reference and stabilize thereafter, indicating that $30$ Hamiltonians per system size are sufficient for near-oracle generalization in this setting.}
  \label{fig:sample_sweep}
\end{figure}

Figure~\ref{fig:sample_sweep} shows the mean test fidelity gap for each held-out test size $L \in \{16,17,18,19,20\}$ as a function of $n_{\mathrm{ham}}$. At $n_{\mathrm{ham}} = 10$, the model already outperforms \majreg\ for the smaller test sizes, although variance across seeds remains high. By $n_{\mathrm{ham}} = 30$, all five test sizes fall below the \majreg\ reference of $0.00613$, and performance stabilizes thereafter with only marginal further improvement. The gap does not increase with test system size within any fixed $n_{\mathrm{ham}}$, confirming that the size-invariant representation extrapolates uniformly across training-set densities. These results indicate that the method is reasonably data-efficient: roughly $30$ Hamiltonians per system size over the range $L \in \{3,\ldots,14\}$ suffice to reach near-oracle performance on out-of-range systems up to $L = 20$ qubits.

\section{Discussion and Limitations}
\label{sec:discussion}

\subsection{Discussion}

The results demonstrate that a transformer trained entirely on  small systems can reliably select high-fidelity Trotter orderings  for larger systems it has never seen, without evaluating any  candidate fidelity at inference time. The strongest non-learned baseline, \majreg, encodes the rule of picking the per-regime majority class. Our model reduces the mean fidelity gap by more than a factor of five relative to this baseline, which indicates that the improvement comes from per-instance information beyond $(p, r)$. From the Hamiltonian terms and Trotter configuration alone, the model learns to select which candidate ordering will best approximate the time evolution for a given instance, a task that would otherwise require expensive fidelity evaluation over all $24$ candidates. Whether this selection reflects physically meaningful structure, such as Trotter commutator norms or localization properties of the Hamiltonian, remains an open question that we plan to investigate in future work through feature attribution and ablation studies. 

Size extrapolation works because the input representation is size-invariant by construction. No absolute qubit index is passed to the model; all features are either per-term local descriptors or scale-invariant global statistics, and the transformer encoder pools over terms without reference to system size. This input design makes extrapolation possible, and the generalization sweep in Section~\ref{sec:data_eff} confirms that it works in practice. How the learned representations evolve with system size is a question we leave to future work.

\subsection{Limitations and Future Directions}
\label{sec:limitations}

Our experiments focus on a single Hamiltonian family (1D XXZ with a transverse field), a single initial state (N\'eel), two Trotter orders, and two step counts, which allowed us to establish a clean proof of concept. The $24$ structured candidates derived from commutation-graph colorings provide a well-suited candidate space for this setting and contain strong orderings across the instances we study. In this setting, many structured candidates are concentrated in a relatively high-fidelity region, so fidelity differences between candidates are often small. In settings where candidate fidelities are more widely separated, classification errors may carry a larger fidelity cost, and it will be important to study whether learned selection remains reliable. In future work, we plan to expand the candidate space by incorporating additional structured orderings from the literature~\cite{tranter2019ordering} and by exploring Hamiltonian-dependent strategies such as symmetry-aware or coefficient-weighted colorings, which may yield better candidates for a broader range of systems.

Extending the approach to 2D lattices, molecular Hamiltonians, or other structured families is a natural next step. These settings introduce richer commutation structure, different term types, and larger term counts. Generalizing across Hamiltonian families will likely require re-architecting the input embedding and the model itself, for example by incorporating richer Pauli-string encodings, geometry-aware features, or family-specific coloring strategies, so that the learned representations transfer across fundamentally different Hamiltonian structures.

Looking further ahead, the long-term goal is direct generation of orderings over the full $k!$ space, allowing the model to propose orderings outside any fixed structured subspace. This would open the door to AI-guided Trotter ordering that is not tied to a predefined candidate set. Future work may explore autoregressive models or pointer networks that learn to directly generate or rank orderings, trained on paired ordering-fidelity data at scale and potentially capable of discovering orderings that no hand-crafted coloring strategy would produce.

\section{Conclusion}
\label{sec:conclusion}
We present a learned approach to Trotter ordering for one-dimensional XXZ Heisenberg Hamiltonians. The method combines a classical analysis of the commutation graph, which produces $24$ structured candidate orderings per instance, with a transformer encoder trained to select among them. At inference time, the selector uses only Hamiltonian and Trotter-configuration features and does not evaluate candidate fidelities. Trained on systems of up to $14$ qubits and evaluated on held-out systems of $16$ to $20$ qubits, the selector achieves a mean test fidelity gap of $0.00115$ relative to the best of the $24$ structured candidates. We show that generalization begins at $L_{\max}=9$, corresponding to training up to $L=8$ qubits with validation at $L=9$, and that larger training ranges bring performance near oracle with little degradation across larger evaluated sizes, including the final held-out range $L \in \{16,\ldots,20\}$. The size-invariant input representation enables this extrapolation across system sizes. To our knowledge, this is the first learned approach targeted specifically at the Trotter ordering problem. These results provide a strong foundation for AI-guided Trotter ordering and motivate future work on generalization across Hamiltonian families, richer candidate spaces, and direct generation of orderings via autoregressive or pointer-network architectures.

\section*{Acknowledgements}

The research presented in this article was supported by the NNSA's Advanced Simulation and Computing Beyond Moore's Law program at Los Alamos National Laboratory. This material is based upon work supported by the U.S. Department of Energy, Office of Science, National Quantum Information Science Research Centers, Quantum Science Center. This work has been assigned LANL technical report number LA-UR-26-23382.

\bibliographystyle{plainurl} %% for arXiv and for conference submissions
\bibliography{refs}

\end{document}